\documentclass{PoS}
\usepackage{epsfig}
\usepackage{graphicx}
\usepackage{caption}

\title{Measurements of the top-quark properties at CMS}

\ShortTitle{Measurements of the top-quark properties at CMS}

\author{\speaker{Efe YAZGAN}\thanks{On behalf of the CMS Collaboration}\\
        Department of Physics and Astronomy, University of Ghent, Proeftuinstraat 86, B-9000 Ghent, Belgium\\
        E-mail: \email{efe.yazgan@cern.ch}}

\abstract{Measurements of several top-quark properties, obtained from the CMS data collected in 2011 and 2012 at centre-of-mass energies of 7 and 8 TeV  are presented. The results include measurements of the top pair charge asymmetry, the W helicity in top quark decays and the search for anomalous couplings, the top quark charge, and of the $t\overline{t}$ spin correlation. The fraction of top quarks decaying into a W-boson and a b-quark relative to all top quark decays, ${\cal R}={\cal B}(t\rightarrow Wb)/{\cal B}(t\rightarrow Wq$), as well as, the cross sections of $t\overline{t}$ events produced in association with a photon or a W or a Z boson are also presented.}

\FullConference{XXI International Workshop on Deep-Inelastic Scattering and Related Subject -DIS2013,\\
		22-26 April 2013\\
		Marseilles,France}

\begin{document}

\section{Introduction}
The most massive particle, the top quark, has a shorter lifetime than the hadronization timescale. 
 This makes its bare quark properties observable. 
The meaurements presented in this note use the data collected with the CMS \cite{ref:cms} 
detector during 2011 and 2012 data taking periods with proton-proton collisions at $\sqrt{s}=$ 7 and 8 TeV, respectively. 

\section{$t\overline{t}$ Charge Asymmetry}
\begin{figure}
\center
\begin{tabular}{cc}
\begin{minipage}[b]{.5\textwidth}
\centering
	\psfig{figure=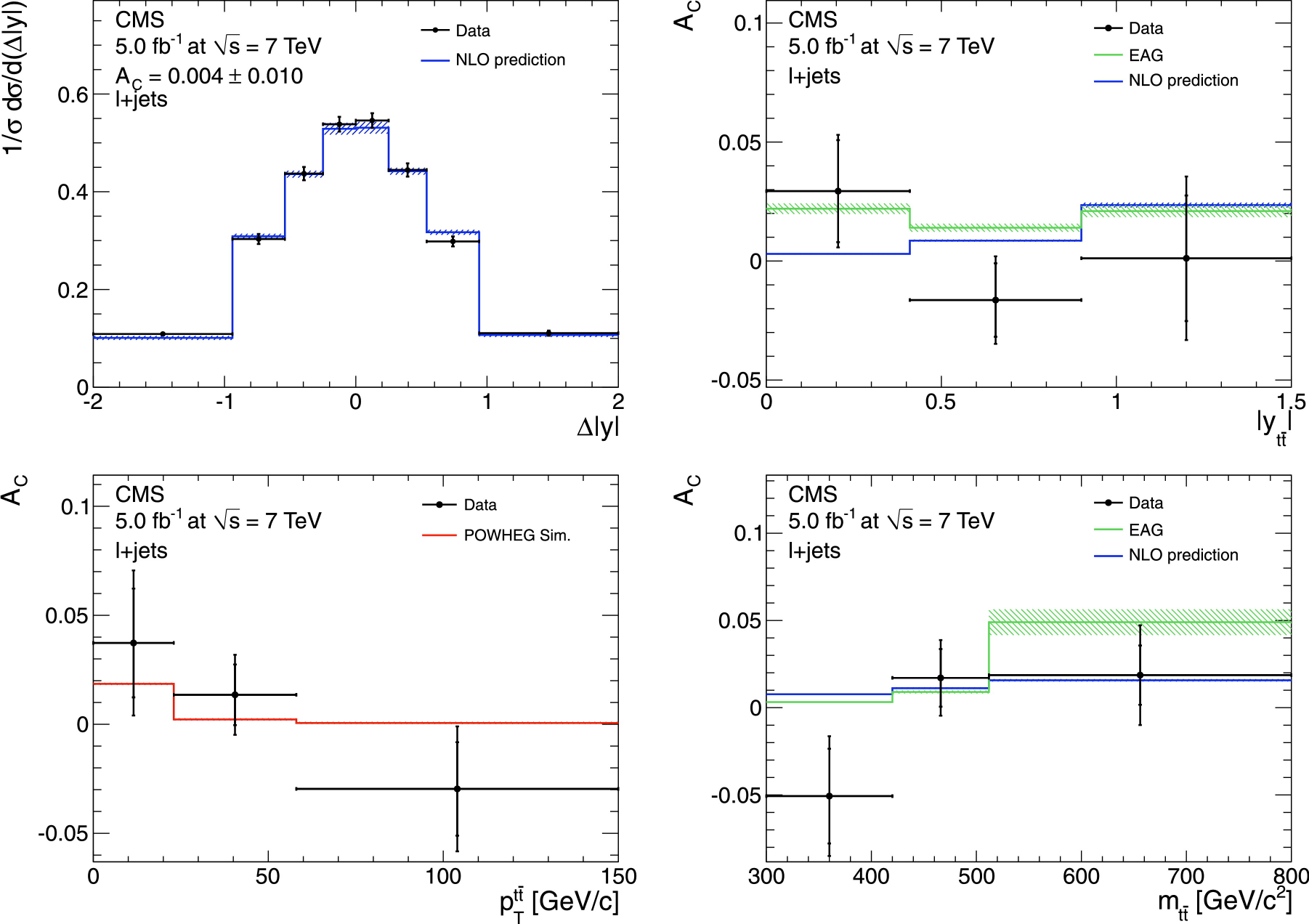, width=1.0\linewidth}
	\caption{Unfolded inclusive $\Delta|y|$ distribution (upper left), $A_C$ vs $|y_{t\overline{t}}|$ (upper right), $p_{t\overline{t}}$ (lower left), and $m_{t\overline{t}}$ (lower right). The measurements 
are compared to SM NLO calculations and to the predictions of an effective model. }
	\label{fig:ac}
\end{minipage}
&
\begin{minipage}[b]{.4\textwidth}
\centering
	\psfig{figure=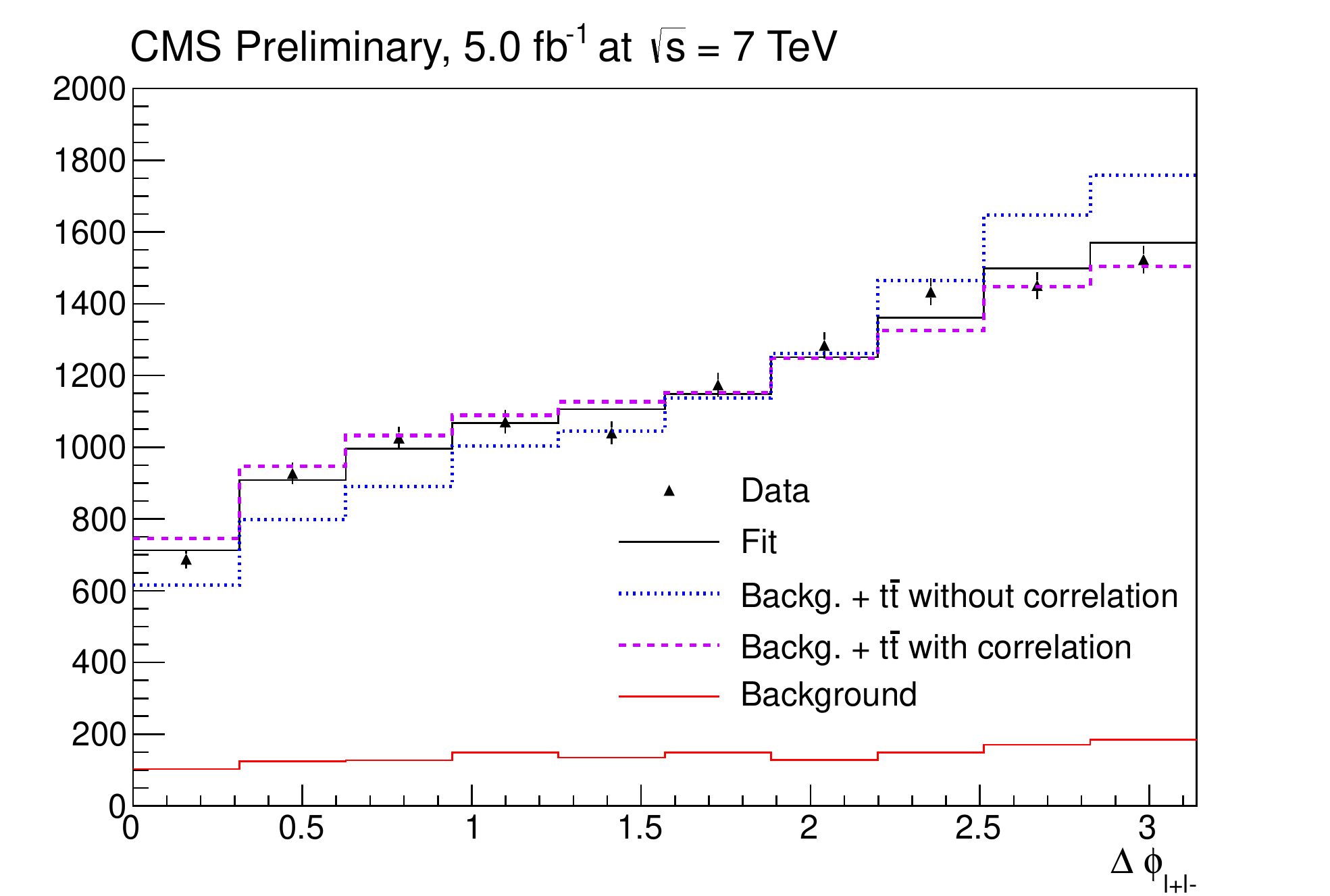, width=1.1\linewidth}
	\caption{Result of the fit (solid line) performed on data (triangles) after the combination of the three channels. The data are also compared to the $\Delta\Phi_{\ell^+\ell^-}$ distribution of $t\overline{t}$ pair events with and without spin correlation. The background components are also shown.}
	\label{fig:sc}
\end{minipage}
\end{tabular}
\end{figure}

Tevatron experiments observed a deviation of $\sim2-3\sigma$ from the standard model (SM) $t\overline{t}$ forward-backward asymmetry \cite{tevAFB1,tevAFB2}. 
The asymmetry occurs in $t\overline{t}$ production through quark-antiquark annihilation. 
The antiquark at the LHC is a sea quark, therefore, on average the quark has a larger momentum fraction.
This results in an excess of top quarks in forward directions. 
Therefore, the  charge asymmetry ($A_C$)  defined using the rapidity difference of the top and the antitop quark ($\Delta|y|=|y_t|-|y_{\overline{t}}|$) is a reasonable choice for the LHC. 
In the lepton+jets channel, the measurement resulted in $A_C=0.004\pm0.010~(stat)\pm0.011~(sys.)$ \cite{AC_ljets} compatible with the SM prediction at NLO, $0.0115\pm0.0006$ \cite{ACnlo1}. 
In the dilepton channel, the inclusive $A_C$ is measured to be 
$0.050\pm0.043~(stat)^{+0.010}_{-0.039}~(sys.)$.  Inclusive $A_C$ is also measured using lepton pseudo-rapidities and is found to be $0.010\pm0.015~(stat)\pm0.006~(sys.)$ \cite{AC_dilepton}. These are in agreement with the NLO predictions of $0.0123\pm0.0005$
and $A_C(\Delta|\eta|)=0.0156\pm0.0007$ \cite{ACnlo2}. Figure \ref{fig:ac} displays ($\Delta|y|$), rapidity ($|y_{t\overline{t}}|$), transverse momenta ($p_{T,t\overline{t}}$), and the invariant mass ($m_{t\overline{t}}$) of the $t\overline{t}$ system.  $|y_{t\overline{t}}|$ depends on the ratio of events with $q\overline{q}$ and $gg$ initial states and at small rapidities, gluon fusion dominates. Therefore, $A_C$ increases with increasing $|y_{t\overline{t}}|$ \cite{ACnlo1} (see Figure \ref{fig:ac} upper right). 
Interferences between the Born and the box diagrams and also between the initial- and final-state-radiation  cause the (anti)top quark direction to be correlated to the initial state (anti)quark. In the former case, there is a positive contribution to $A_C$ and a negative contribution in the latter. 
In the presence of initial- or final-state-radiation the $t\overline{t}$ system has a higher $p_{T}$, hence there is more negative contribution to $A_C$ for such events \cite{ACnlo1} (see Figure \ref{fig:ac} lower left). 
At high invariant mass of the $t\overline{t}$ system, $m_{t\overline{t}}$, number of $t\overline{t}$ events with $q\overline{q}$ production increases and therefore $A_C$ increases (see Figure \ref{fig:ac} lower right).  However, if new heavy particles have a role in $t\overline{t}$ production, they can interfere with the SM processes and this might result in a different dependence on $m_{t\overline{t}}$. 
The distributions in Figure \ref{fig:ac}  are also compared to the distributions predicted by a model with an effective axial-vector coupling of the gluon \cite{eft1,eft2} that could describe the depence of forward-backward asymmetry vs $m_{t\overline{t}}$  observed at the Tevatron. None of the distributions in Figure \ref{fig:ac} shows a deviation from the SM expectations however the uncertainties are still large. 

\section{W Boson Helicities in Top Quark Decays and Anomalous Couplings}
Partial witdh of the top quark can be parametrized in terms of left-handed (F$_L$), right-handed ($F_R$), longitudinal ($F_0$) W boson helity fractions and 
the $\theta^*$ angle between the momentum of the d-type fermion in the $W$ rest frame and  of the $W$ in the top quark rest frame. 
$W$ helicities in $t\bar{t}$ events in dilepton and lepton+jets final states are measured \cite{cmswhel1,cmswhel2} . Measurements at $\sqrt{s}=$ 7 TeV made by CMS and ATLAS are also combined \cite{ref:whelcomb}. 
Individual and combined measurements are found to be consistent with each other and with SM predictions at NNLO QCD. 
Measurement of W boson helicity fractions is also performed in single-top topologies \cite{ref:whelsingletop} in the $\mu+jets$ final state at $\sqrt{s}$ = 7 and 8 TeV. The helicity fractions are obtained from likelihoods with reweighted signals including all processes involving the top quark. 
The measurements performed at $\sqrt{s}$ = 7 and 8 TeV are combined yielding $F_0=0.713\pm0.114$ (stat)$\pm0.023$ (syst), $F_L=0.293\pm0.069$ (stat)$\pm0.030$ (syst) and $F_R=-0.006\pm0.057$ (stat)$\pm0.027$ (syst).  The measurements are consistent with the SM predictions and the measurements in the $t\bar{t}$ channels.
These measurements are also used to set limits on the real part of the anomalous $Wtb$ couplings. 

\section{Top Quark Charge}
Being the electroweak isospin partner of the b-quark, top quark has an electric charge of $+2/3e$. 
A measurement is made testing two different top quark charge hypotheses in the muon+jets final state at $\sqrt{s}=7$ TeV \cite{cmstopcharge}. 
For the measurement, charge correlations between muons from W boson decays and soft muons from B-hadron decays in b-jets are utilized to constrain the top quark charge. Using a normalized asymmetry variable,  the top quark charge hypothesis of $-4/3e$ against $+2/3e$ is tested. The measurement yielded an asymmetry of 
$A=0.97\pm0.12~(stat.)\pm0.31~(sys.)$ in agreement with the SM expectation of $A=+1$. 

\section{$t$-$\overline{t}$ Spin Correlation}
The spin-decorrelation timescale, $m_t/\Lambda_{QCD}^2\sim3\times10^{-21}$ s, is larger than the hadronization time scale. 
Therefore spin effects propagate to the decay products. In the dilepton final state, the  azimuthal angle between the opposite charge leptons, $|\Delta\phi_{\ell^{+}\ell^{-}}|$, in the laboratory frame is sensitive to the $t$-$\overline{t}$ spin correlation. 
This variable can be measured precisely without top quark reconstruction. 
At $\sqrt{s}=7$ TeV, the $t$-$\overline{t}$ spin correlation in the dilepton channel is extracted from a template fit to the $|\Delta\phi_{\ell^{+}\ell^{-}}|$ distribution \cite{SC}.   Three different templates are used; simulated $t\overline{t}$ events assuming the SM, simulated $t\overline{t}$ events without spin correlation and a template for the background events. 
A binned likelihood fit to data is used to simultaneously fit the $e^+e^-$, $\mu^+\mu^-$ and $e^{\pm}\mu^{\mp}$ channels to determine the fraction of events with SM spin correlation, $f=N_{SM}/(N_{SM}+N_{uncor})$. 
The measurement yielded  $f=0.74\pm0.08~(stat.)\pm0.24~(sys.)$ consistent with the SM expectation of $f^{SM}=1$. The result of this fit performed on data after the combination of the three channels is shown in Figure \ref{fig:sc}. The data are also compared to the $|\Delta\phi_{\ell^{+}\ell^{-}}|$ distributions obtain from $t\overline{t}$ events with and without spin-correlation. 
The fraction of events with SM spin correlation, $f$, can be converted to a measurement of the spin correlation coefficient $A$. 
In the helicity basis, the measurement yields $A_{hel}^{meas}=0.24\pm0.02~(stat.)\pm0.08~(sys.)$ consistent with the SM prediction at NLO of $A_{hel}^{SM}=0.31$ \cite{SCbern}. 
The spin correlation is also studied using other variables and at different $m_{t\overline{t}}$ values \cite{SC}. 
All measurements are found to be consistent with the SM predictions. 

\section{Measurement of the Ratio ${\cal R}={\cal B}(t\rightarrow Wb)/{\cal B}(t\rightarrow Wq$)}
The top quark decays to a W boson and a b quark practically with a branching ratio of $\sim100$\%.
Decays to other down-type quarks are suppressed in the CKM matrix.  
A measurement of $|V_{tb}|$ is made by measuring the branching fraction ratio ${\cal R}={\cal B}(t\rightarrow Wb)/{\cal B}(t\rightarrow Wq$)
in the $t\overline{t}$ di-lepton final state using $\sqrt{s}=$8 TeV data \cite{ref:rmeas}. 
${\cal R}$ is extracted from a profile likelihood fit to data-driven analytic probability models of signal purity, number of reconstructed tops in different jet categories and number of b-tags. The fit gave ${\cal R}=1.023^{+0.036}_{-0.034}$ (stat+syst).
Imposing ${\cal R}\leq1$, a lower limit for ${\cal R}$ is derived to be ${\cal R}>0.945$ at 95\% C.L. 
With the assumption of CKM unitarity and existence of three generations, the measurement is converted to a measurement of $|V_{tb}|$ yielding $1.011^{+0.018}_{-0.017}$ and a lower limit is derived to be $|V_{tb}|>0.972$ at 95\% C.L.  The results obtained for ${\cal R}$ and $|V_{tb}|$ are consistent with the SM predictions and represent the most precise measurement of ${\cal R}$ and the most stringent direct limit on $|V_{tb}|$. 


\section{Vector Boson Production Associated with Top-Antitop Pairs}
Measurement of vector boson production in association with $t\bar{t}$ pairs provides a test of the SM top quark-vector boson couplings. Measurement of these couplings are important for new physics searches and also the measurements of the Higgs boson in the $t\bar{t}H$ process. The cross section measurements of the $t\bar{t}V$ processes are made  using the same-sign dilepton for $t\bar{t}W$ and $t\bar{t}Z$ and trilepton signature for $t\bar{t}Z$ process at $\sqrt{s}=7$ TeV \cite{ref:ttvcms}. The cross section for $t\bar{t}V$ is measured  to be $0.43^{+0.17}_{-0.15} (stat)^{+0.09}_{-0.07} (syst)$ pb and $t\bar{t}Z$ to be $0.28^{+0.14}_{-0.11} (stat)^{+0.06}_{-0.03} (syst)$ pb consistent with the SM NLO calculations \cite{ref:cambell12,ref:garzelli12}.
The $t\bar{t}Z$ measurements represent the first direct measurements of the top quark-Z boson coupling. 

\section{Conclusions}
Measurements of top quark properties are providing thorough tests of the SM. 
All top quark properties measurements at the LHC are in good agreement with the SM predictions.


\begin{thebibliography}{99}
\bibitem{ref:cms} CMS Collaboration, \emph{The CMS experiment at the CERN LHC}, \emph{JINST}  {\bf 3} (2008) S08004.
\bibitem{tevAFB1} CDF Collaboration, \emph{Measurement of the top quark forward-backward production asymmetry and its dependence on event kinematic properties}, \emph{PRD} {\bf 87} (2013) 092002 [arXiv:1211.1003 [hep-ex]]. 
\bibitem{tevAFB2} D0 Collaboration, \emph{Forward-backward asymmetry in top quark-antiquark production}, \emph{PRD} {\bf 84} (2011) 112005 [arXiv:1107.4995 [hep-ex]].
\bibitem{AC_ljets} CMS Collaboration, \emph{Inclusive and differential measurements of the $t\overline{t}$ charge asymmetry in proton$-$proton collisions at $\sqrt{s}=$7 TeV},  \emph{PLB} {\bf 717} (2012) 129 [arXiv:1207.0065 [hep-ex]].
\bibitem{ACnlo1} J. H. Kuhn, and G. Rodrigo, \emph{Charge Asymmetries of top quarks at hadron colliders revisited}, \emph{JHEP} {\bf 01} (2012) 063 [arXiv:1109.6830 [hep-ph]]. 
\bibitem{AC_dilepton} CMS Collaboration, \emph{Top charge asymmetry measurement in dileptons at 7 TeV}, \emph{CMS-PAS-TOP-12-010} (2012).  
\bibitem{ACnlo2} W. Bernreuther and Z-G. Si, \emph{Top quark and leptonic charge asymmetries for the Tevatron and LHC}, \emph{PRD} {\bf 86} (2012) 034026 [arXiv:1205.6580 [hep-ph]]. 
\bibitem{eft1} E. Gabrielli, A. Racioppi, M. Raidal, \emph{Implications of the effective axial-vector coupling of the gluon on top$-$quark charge asymmetry at the LHC}, \emph{PRD} {\bf 85} (2012) 074021 [arXiv:1112.5885v1 [hep-ph]].  
\bibitem{eft2} G. Brooijmans, et al., \emph{Les Houches 2011: Physics at TeV colliders new physics
working group report}, (2012) [arXiv:1203.1488v2 [hep-ph]].
\bibitem{cmswhel1} CMS Collaboration, \emph{Measurement of the W boson polarization in semileptonic top pair decays with the CMS detector at the LHC}, \emph{CMS-PAS-TOP-11-020} (2011).
\bibitem{cmswhel2} CMS Collaboration, \emph{Measurement of the W helicity in top pair production with dileptons at 7 TeV}, \emph{CMS-PAS-TOP-12-015} (2012).
\bibitem{ref:whelcomb} ATLAS and CMS Collaborations, \emph{Combination of the ATLAS and CMS measurements of the W-boson polarization in top-quark decays}, \emph{ATLAS-CONF-2013-033, CMS-PAS-TOP-12-025} (2013).
\bibitem{ref:whelsingletop} CMS Collaboration, \emph{W-helicity measurement in single top events topology}, \emph{CMS-PAS-TOP-12-020} (2012).
\bibitem{cmstopcharge} CMS Collaboration, \emph{Constraints on the Top-Quark Charge from Top-Pair Events}, \emph{CMS-PAS-TOP-11-031} (2011). 
\bibitem{SC} CMS Collaboration, \emph{Measurement of Spin Correlations in $t\overline{t}$ events in the dilepton channels in pp collisions at $\sqrt{s}$ = 7 TeV}, \emph{CMS-PAS-TOP-12-004} (2012). 
\bibitem{SCbern} W. Bernreuther and Z.-G. Si, \emph{Distributions and correlations for top quark pair production and decay at the Tevatron and LHC}, \emph{Nuc. Phys. B} {\bf 837} (2010) 90 [arXiv:1003.3926 [hep-ph]].
\bibitem{ref:rmeas} CMS Collaboration, \emph{Measurement of the ratio $B(t\rightarrow Wb)/B(t\rightarrow Wq)$}, \emph{CMS-PAS-TOP-12-035} (2012). 
\bibitem{ref:ttvcms} CMS Collaboration, \emph{Measurement of associated production of vector bosons and top quark-antiquark pairs in pp collisions at $\sqrt{s}=$7 TeV}, \emph{PRL} 110 (2013) 172002   [arXiv:1303.3239 [hep-ex]].
\bibitem{ref:cambell12} J. M. Campbell and R. K. Ellis, \emph{$t\overline{t}W^{\pm}$ production and decay at NLO}, \emph{JHEP} {\bf 07} (2012) 052.
\bibitem{ref:garzelli12} M.V. Garzelli, A. Kardos, C.G. Papadopoulos, Z. Trocsanyi \emph{$t\overline{t}W^{\pm}$ and $t\overline{t}Z$ Hadroproduction at NLO accuracy in QCD with Parton Shower and Hadronization effect},  \emph{JHEP} {\bf 11} (2012) 056.

\end{thebibliography}
\end{document}